\documentclass[a4paper,11pt]{article}

\usepackage{jcappub}

\usepackage[T1]{fontenc}

\usepackage{graphicx}%
\usepackage{dcolumn}
\usepackage{amsmath}
\usepackage{bm}

\title{\boldmath Aspects of Wave Turbulence in Preheating}

\author[a]{Jos\'e A. Crespo}
\author[a,1]{and H. P. de Oliveira\note{Corresponding author.}}

\affiliation[a]{{\it Universidade do Estado do Rio de Janeiro }\\
{\it Instituto de F\'{\i}sica - Departamento de F\'{\i}sica Te\'orica,}\\
{\it CEP 20550-013. Rio de Janeiro, RJ, Brazil.}}

\emailAdd{jaacrespo@gmail.com}
\emailAdd{oliveira@dft.if.uerj.br}

\abstract{In this work we have studied the nonlinear preheating dynamics of the $\frac{1}{4} \lambda \phi^4$ inflationary model. It is well established that after a linear stage of preheating characterized by the parametric resonance, the nonlinear dynamics becomes relevant driving the system towards turbulence. Wave turbulence is the appropriated description of this phase since matter distributions are fields instead of usual fluids. Therefore, turbulence develops due to the nonlinear interations of waves, here represented by the small inhomogeneities of the inflaton field. We present relevant aspects of wave turbulence such as the Kolmogorov-Zakharov spectrum in frequency and wave number domains that indicates that there are a transfer of energy through scales. From the power spectrum of the matter energy density we were able to estimate the temperature of the thermalized system.}

\begin{document}
\maketitle
\flushbottom

\section{Introduction}

One of the most significant challenges in modern Cosmology is the description of the early stages of the universe. There is a general acceptance that an inflationary phase \cite{inflation} characterized by a huge expansion of the universe might have occurred that preceded the radiation dominant phase. In this context,  the process of reheating plays a crucial role in the transition of the Universe from the inflationary phase into the radiation phase, and consequently in the creation of almost all matter constituting the present Universe.

The reheating begins at the end of inflation with a stage of parametric resonance with a rapid transfer of energy from the inflaton field into other matter fields, leading to particle production and the inflaton decay, far away from thermal equilibrium. Several authors have explored \cite{linde,linde1,kleb,kaiser,deol1,deol2,review,anderson} the entire evolution of the inflaton field until the universe has settled down in a  thermalized state characterizing the radiation era. The main feature shared by these studies on the later stages of non-perturbative preheating is the ubiquity of turbulence in the process towards thermalization. However, the matter constituents at the end of inflation are fields together with their small inhomogeneities rather conventional fluids, which makes more appropriate to deal with \textit{wave turbulence} \cite{nazarenko} instead hydrodynamic turbulence \cite{frish}.

An appropriate and useful definition of wave turbulence is a state of out-of-equilibrium statistical mechanics of random nonlinear waves \cite{nazarenko}. The most important class of solutions in wave turbulence are called Kolmogorov-Zakharov (KZ) spectra \cite{zakharov} that correspond to a constant flux of energy through scales, where scaling laws spectra in frequency and wave numbers are present. In the context of preheating the small inhomogeneities associated to the fields play the role of the waves, some of them are resonant in the first phase of preheating. The nonlinear interaction of these waves results in the transfer of energy from the homogeneous inflaton field through different scales with the establishment of turbulence in this scenario.
  
In this paper,  our main goal is to explore aspects of wave turbulence of late stages of preheating exhibiting the KZ spectrum in frequency and wave number. We have considered a single field inflationary model with quartic potential $\frac{1}{4} \lambda \phi^4$, where $\phi$ is the inflaton field. This is the simplest model with a self interacting inflaton field and also the first model in which the nonlinear effects of preheating were studied. Although the observational data have ruled out this class of inflationary model \cite{planck}, we claim that the features of wave turbulence are robust and present in more realistic models. We have organized the paper as follows. We introduce the basic equations of the model and the numerical treatment based on spectral methods in the second Section. In Section 3,  we present the numerical results starting from a standard verification of the accuracy of the numerical method. In the sequence, we present the relevant aspects of wave turbulence that are the scaling laws associated to the relevant quantities in frequency and wave number. In Section 4,  we briefly discussed the effect of backreaction and the evolution of the equation of state. Finally, in Section 5 we conclude.

\section{The model}%

Let us consider the simplest model of one field preheating that has already been studied  in several works \cite{micha,deol1,deol2}. We denote $\phi({\bf x},t)$ as the inflaton field with potential $V(\phi)=\frac{1}{4}\lambda{\phi}^4$ evolving in a spatial flat Friedmann-Robertson-Walker universe \cite{linde,linde1}. The homogeneous component, or simply the homogeneous mode of the inflaton  $\phi_0(t)$, is responsible by the inflationary. At the end of inflation, we donote $\phi_0(0)=\phi_e$ the homogeneous mode at the end of inflation. It will be useful to introduce the conformal time $\tau$ defined by $a(\tau)d\tau=\sqrt{\lambda} \phi_{0}(0)a(0) d t$, and the conformal scalar field $\varphi({\mathbf{x}},\tau)=\phi a(\tau)/\phi_{0}(0) a(0)$, with $\tau$ being the scale factor. The evolution equation for the inflaton is written in a simple and convenient way:

\begin{equation}
\varphi^{\prime\prime}-\nabla^2 \varphi-\frac{a^{\prime\prime}}{a}\varphi+{\varphi}^3=0,
\label{eq1}
\end{equation}

\noindent where prime indicates derivative with respect to the conformal time $\tau$, with $\tau=0$ signalizing the end of inflation. The beginning of preheating is characterized by coherent oscillations of the inflaton which yields an effective traceless energy-momentum tensor \cite{turner}. As a consequence, $a(\tau) \sim \tau$, therefore, allowing us to set $a^{\prime\prime}=0$ in Eq. (\ref{eq1}).

We have integrated numerically Eq. (\ref{eq1}) using the collocation or pseudospectral method \cite{collocation} in a two dimensional square box of size $L$ with periodic boundary conditions, but there are some other numerical approaches applied to this problem \cite{varios_num}. As in any spectral method, we have approximated the conformal scalar field is approximated as a series with respect to a set of basis functions. According to the boundary conditions, these basis function are Fourier functions, and the spectral approximation establishes that,

\begin{eqnarray}
\varphi({\bf x},t) = \sum_{l,m=-N}^N a_{l m}(\tau) \psi_{l m}(x,y).
\label{eq2}
\end{eqnarray}
 
\noindent In this expression $N$ is the truncation order that limit the number of unknown modes $a_{lm}(\tau)$, the basis function are $\psi_{{\bf k}}=\exp\left(\frac{2 \pi i}{L}\,{\bf k.x}\right)$, and ${\bf k} = (l,m)$ is the comoving momentum. Notice that the mode $a_{\mathbf{0}}(\tau)$ corresponds to the homogeneous component of the inflaton while the remaining modes account for its small inhomogeneities.

We have used the standard Galerkin method to solve this problem previously \cite{deol1,deol2}, but with low truncation order once the equations were expressed solely in terms of the modes. The collocation method adopts a more convenient approach to deal with the nonlinearities by establishing that the residual equation - the equation arising when the spectral approximation (\ref{eq2}) is substituted into the evolution equation (\ref{eq1}) - vanishes at particular points named collocation or grid points $\mathbf{x}_{\mathbf{k}}=2\pi\mathbf{k}/(2N+1)$. As a consequence, we have $\mathrm{Res}(\mathbf{x}_\mathbf{k},\tau) = 0$, with $\mathbf{k}=(l,m)$ and $l,m=1,2,..,2N+1$, and the resulting equations are expressed schematically by, 
 
\begin{equation}
\varphi^{\prime\prime}_{{\bf k}}(\tau) + \sum_{\bf{j}} \omega^2_{{\bf j}} a_{{\bf j}}(\tau)\psi_{\bf j}(x_{\bf k}) +
\varphi^3_{{\bf k}}(\tau) = 0,
\label{eq3}
\end{equation}

\noindent where $\omega^2_{\bf j} = \frac{4 \pi^2}{L^2}\,{\bf j}^2$ and $\varphi_{\mathbf{k}}(\tau)$ denotes the value of the scalar field at the collocation point $\mathbf{x_k}$. It becomes very economical to express the resulting equations using both representations of the scalar field: the spectral representation through the modes $a_{\bf{k}}$, and the physical representation through values of the scalar field at the collocation points. There are $(2N+1)^2$ independent equations that coincide with the same number of independent modes $a_{\bf{k}}(\tau)$ recalling that these modes are decomposed into imaginary and real pieces, but not all are independent on the grounds of the scalar field given by Eq. (\ref{eq1}) is real \cite{deol1}. The bridge connecting both representation consists in the following relation,

\begin{equation}
\varphi_{\bf{k}}(\tau) = \sum_{l,m=-N}^N\,a_{lm}(\tau) \psi_{lm}(\mathbf{x_k}).
\label{eq4}
\end{equation} 

\noindent Accordingly, the $(2N+1)^2$ values $\varphi_{\bf{k}}(\tau)$ are related to an equal number of independent modes $a_{\bf j}(\tau)$. In all numerical simulations we have evolved the dynamical equations (\ref{eq3}) with a fourth-order Runge-Kuta integrator.

\section{Numerical results}

We have integrated the dynamical equations (\ref{eq3}) with the initial conditions $(\varphi_{\bf{k}}(0),\varphi^\prime_{\bf{k}}(0))$ at the end of inflation. In accordance to the choice of the conformal scalar field, its homogeneous component is set initially as $a_{\bf{0}}(0)=1$, whereas the other modes, $a_{\bf{j}}(0)$, have amplitudes of order of $10^{-4}$. With respect to the associated velocities, we have $a^\prime_{\bf{0}}(0)=0$ and $|a^\prime_{\bf{j}}(0)| \sim \mathcal{O}(10^{-4})$ (for more details see Refs. \cite{micha,deol1,deol2,kleb}). 

\begin{figure}[htb]
\centering
\includegraphics[height=6.5cm,width=8.5cm]{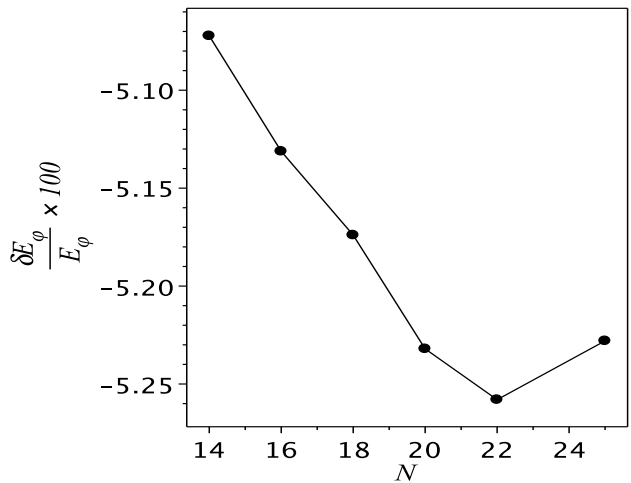}
\caption{Log-linear plot of the average error $\delta E_\varphi/E_\varphi \times 100$ over the interval from $\tau=0$ to $\tau=1,100$ for $\lambda=10^{-4}$. The error exponential decay of the error is a confirmation of the accuracy of the numerical method.}
\end{figure}

\begin{figure}[htb]
\centering
\includegraphics[height=7.5cm,width=7.5cm]{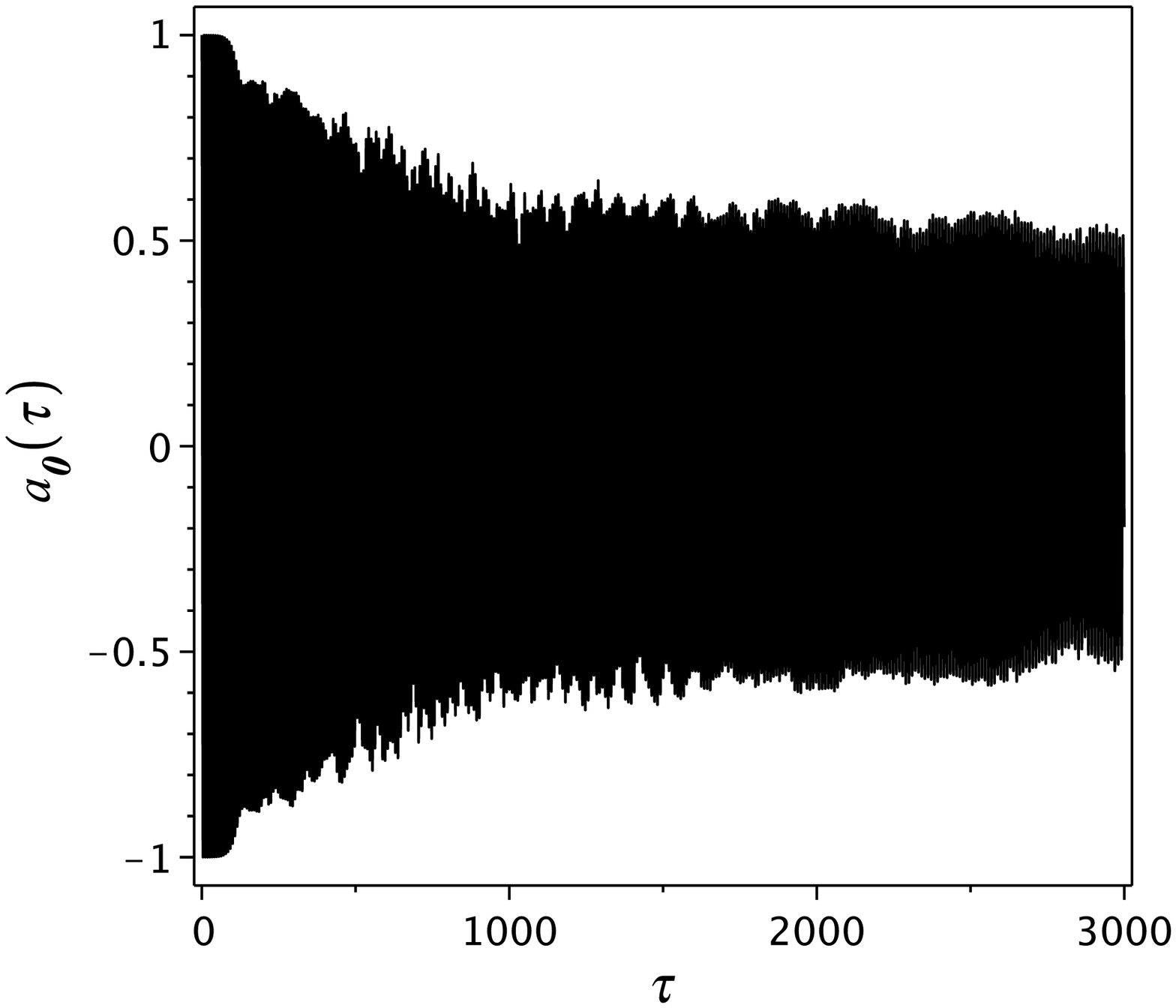}
\includegraphics[height=7.5cm,width=7.5cm]{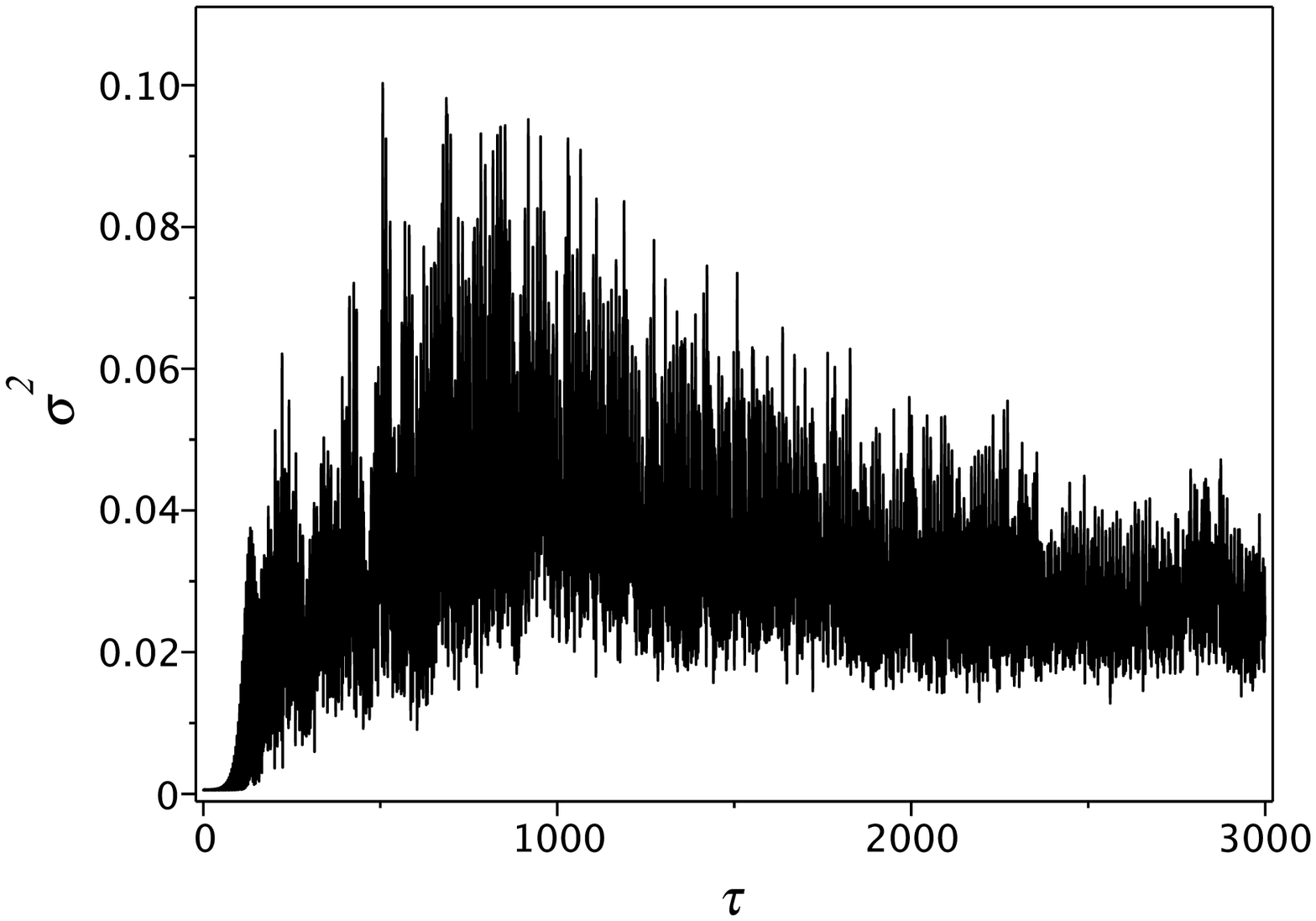}
\caption{Evolution of the homogeneous component of the inflaton field $a_{\mathbf{0}}(\tau)$ and the spatial average of the variance $\sigma^2(\tau)$. The qualitative behavior is in agreement with other similar plots, and also for distinct models. Here we have set $\lambda=10^{-4}$ so that the homogeneous mode starts to evolve in an approximately constant amplitude after $\tau \approx 1,300$.}
\end{figure}

The parameter $L$ dictates which modes with wave vector $\mathbf{k}=(l,m)$ undergo an initial phase of parametric resonance by considering the stability/instability chart for the Lam\'e equation that governs the evolution of the modes $a_{\mathbf{j}}(\tau)$ in the linearized regime \cite{linde,deol1}. In this case, it can be shown that if $L = \pi \sqrt{5 (l^2+m^2)/2}$, where $l,m$ assumes any integer value in the interval $[-N,..,N]$, then those modes with $|\mathbf{k}|=\sqrt{l^2+m^2}$ will grow exponentially in the first stages of preheating. In our numerical simulations, we have chosen these resonant modes with $|\mathbf{k}|=\sqrt{5}$, or equivalently $L = 5 \pi/\sqrt{2}$.

Before proceeding with our numerical study, we present a numerical test that consists in verifying the conservation of the total energy of the conformal inflaton field, $E_{\varphi}(\tau)$, 

\begin{equation}
E_\varphi(\tau) = \int_{\mathcal{D}} \rho_\varphi(\tau,\mathbf{x}) d^2 \mathbf{x}, \label{eq5}
\end{equation}

\noindent where $\mathcal{D}$ represents the spatial domain and $\rho_\varphi = 1/2 \varphi^{\prime 2} + 1/2 (\nabla \varphi)^2 + 1/4 \varphi^4$. As a matter of fact, this is valid only if $a^{\prime\prime} = 0$ in Eq. \ref{eq1}. A very useful way of checking the energy conservation is to evolve the relative variation of energy given by $\delta E_\varphi(\tau)/E_\varphi(0) \times 100$ with several truncation orders and evaluate the root mean square deviation for each truncation order. In Fig. 1, we present the results that show a good convergence to a relative deviation of about $10^{-5}$. 

The main stages of the dynamics of the inflaton have been described with details in Refs. \cite{micha,deol1,deol2}. We do not intend to repeat the description of these stages towards turbulence, but illustrate them by displaying in Fig. 2 the long time behavior of the homogeneous component of the inflaton or the homogeneous mode, $a_{\mathbf{0}}(\tau)$, and $\sigma^2(\tau)$ defined by,

\begin{equation}
\sigma^2(\tau) = \left<(\varphi-\left<\varphi \right>)^2 \right> = \sum_{\mathbf{k}}\,|a_\mathbf{k}|^2-a_\mathbf{0}^2, \label{eq6}
\end{equation}

\noindent where $\mathrm{var}(\varphi) = (\varphi-\left<\varphi \right>)^2$ is the variance of the field $\varphi$, and $\left<..\right> = 1/L^2\,\int_{\mathcal{D}} .. d^2 \mathbf{x}$ is the average over the spatial domain. Using expression (\ref{eq4}) it follows that homogeneous mode is the spatial average of the inflaton field $\left<\varphi \right>=a_{\mathbf{0}}$.

In the numerical experiments, we have set $\lambda=10^{-4}$ and $\lambda=10^{-8}$. In both cases, the structure of the time signals of $a_{\mathbf{0}}(\tau)$ and $\sigma^2(\tau)$ is essentially the same. The only difference is the time scale that separates approximately the stages until the turbulent phase establishes. In particular, we call attention to the decay of the homogeneous mode amplitude which starts when the resonant modes grow beyond the linear regime. We have verified that the amplitude decays approximately as $\tau^{-1/3}$ in agreement with previous studies \cite{kleb} until a certain time, beyond which the amplitude remains approximately constant. For the homogeneous mode of Fig. 2, this occurs at $\tau \approx 1,300$ and, for $\lambda= 10^{-8}$, we found $\tau \approx 14,000$. We interpret this stage as representing the regime of stationary turbulence identifying it as the thermalized phase. In what follows we have considered this phase to exhibit some aspects of wave turbulence.

One of the most relevant features of any turbulent signal concerns, not to its detailed structure, but to some property that is reproducible and related to the statistical description of turbulence. We exhibit this property by constructing a sequence of histograms of $\sigma^2(\tau)$ considering intervals of time, $\Delta \tau=100,300,600,800$, about $\tau=2,350$. As shown in Fig. 3 the histograms are identical, and the same axisymmetric Gaussian distribution fits them.

\begin{figure*}[htb]
\centering
\includegraphics[scale=0.3]{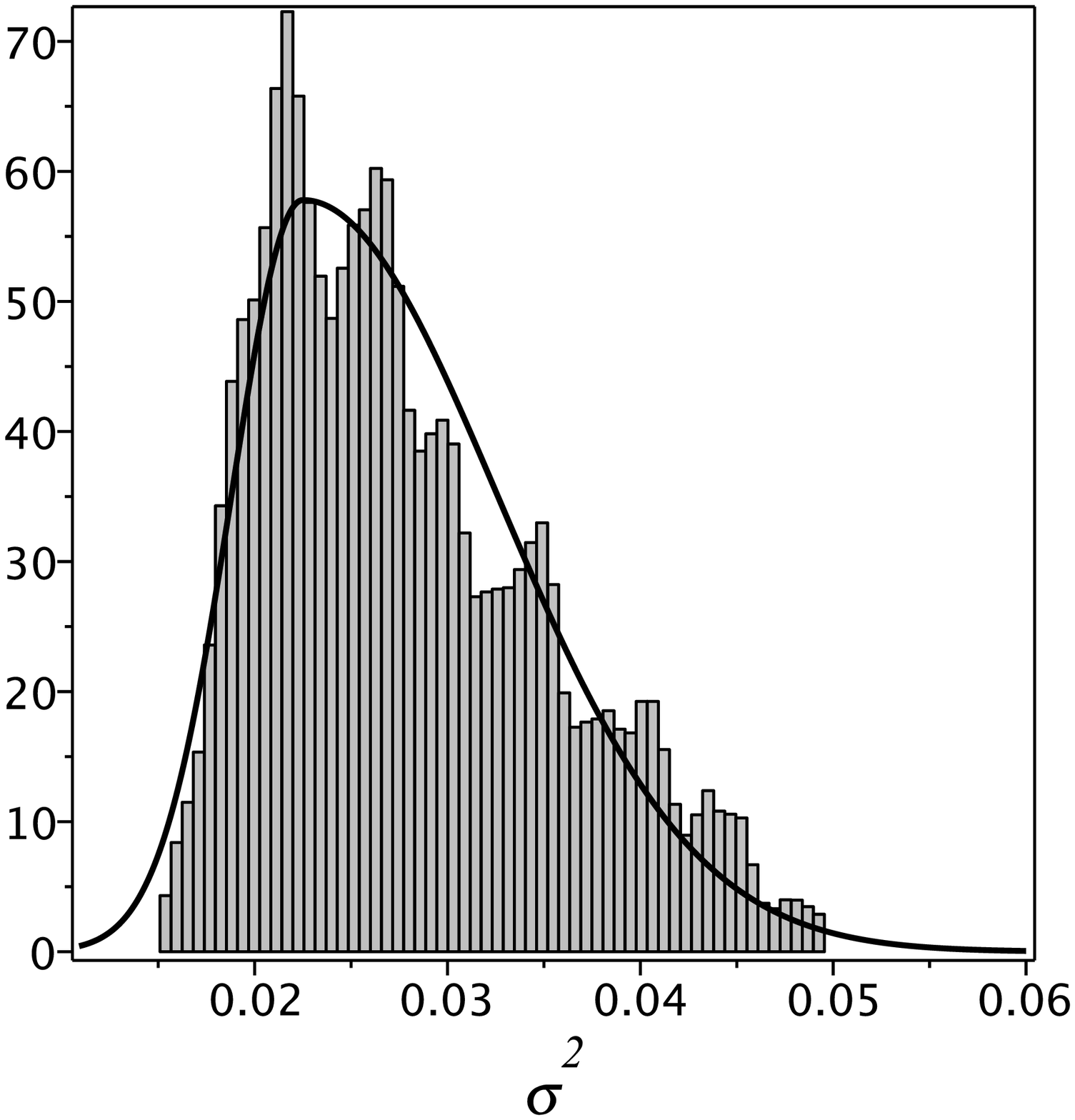}\includegraphics[scale=0.3]{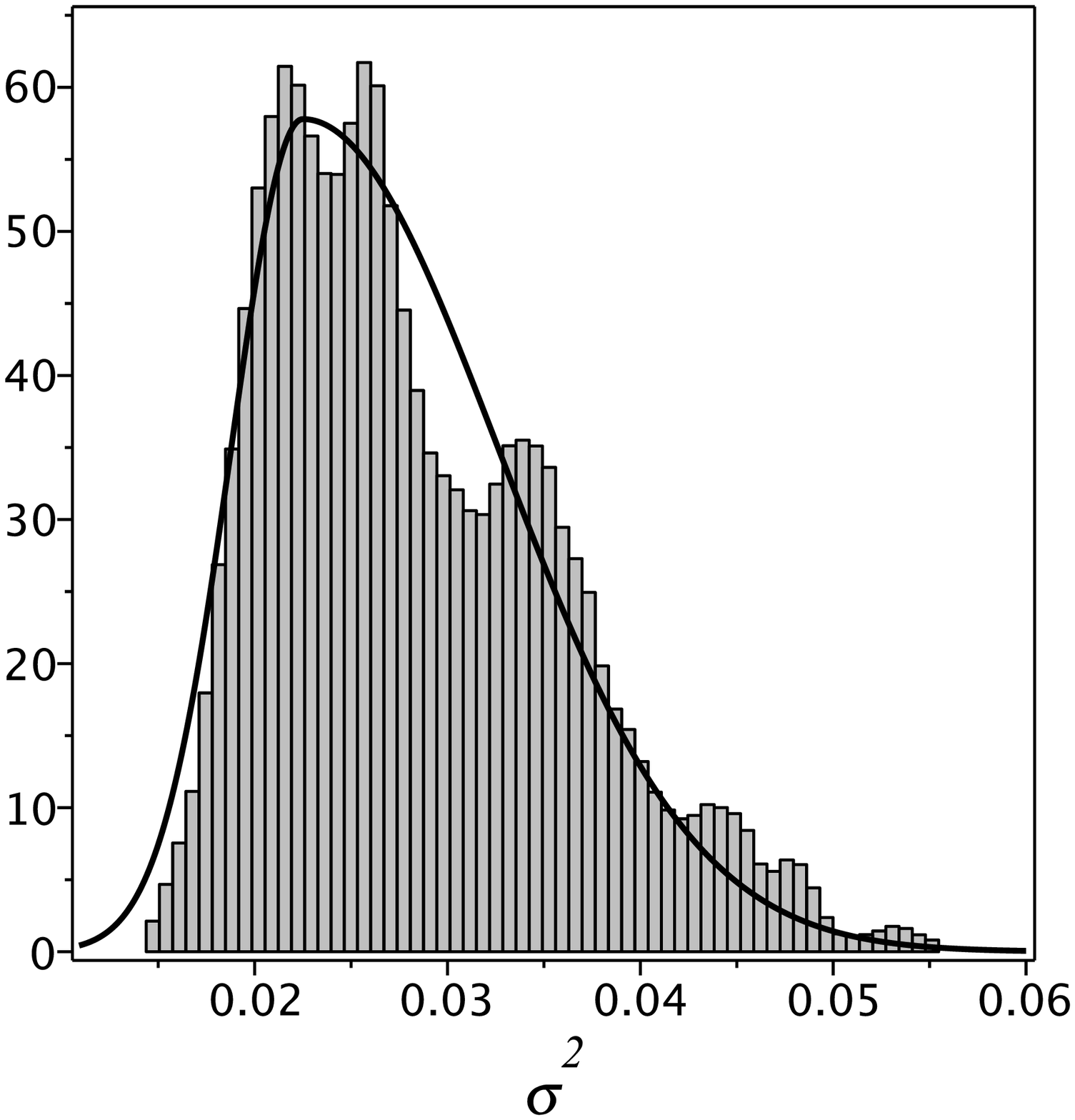}
\includegraphics[scale=0.3]{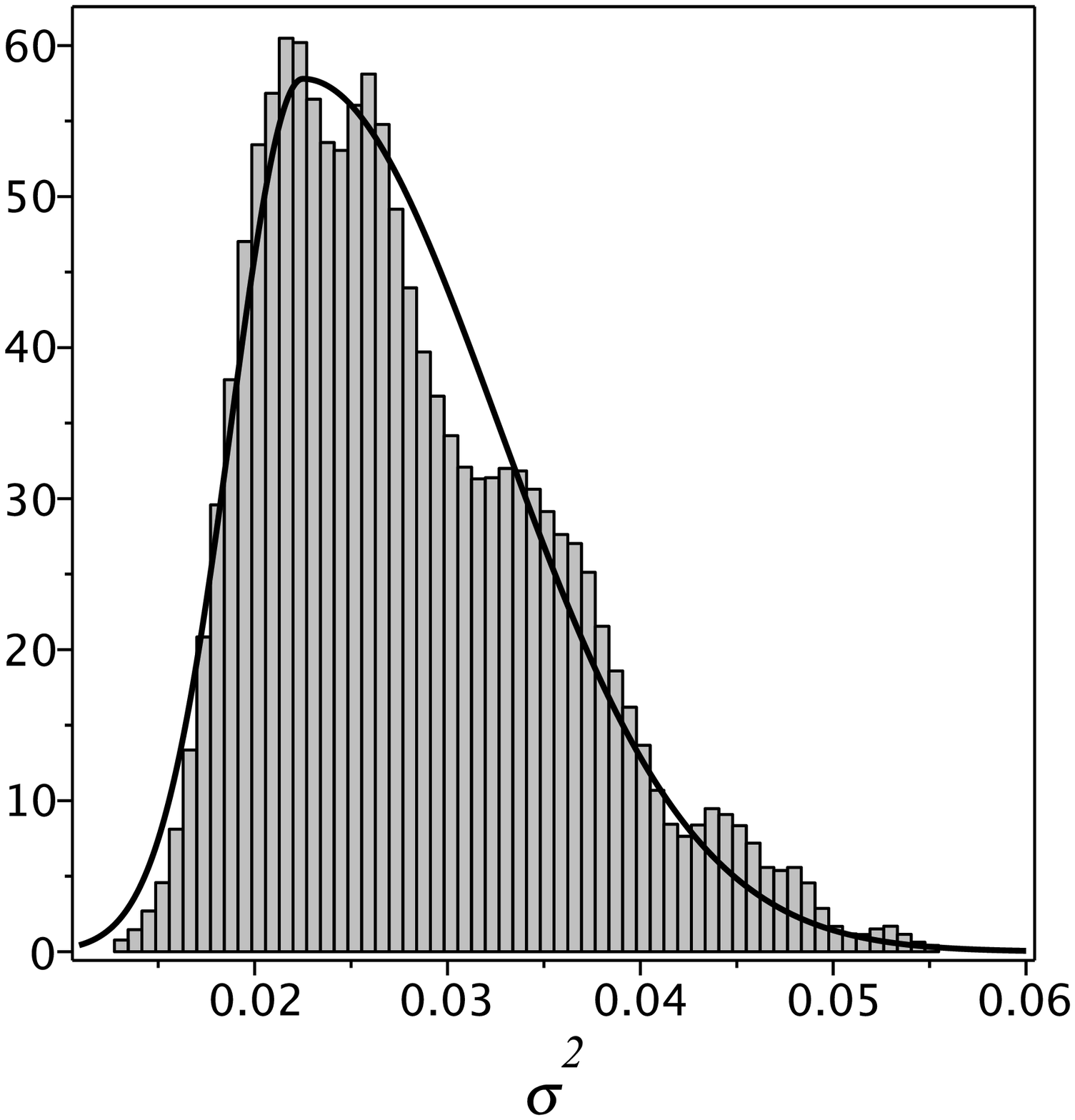}\includegraphics[scale=0.3]{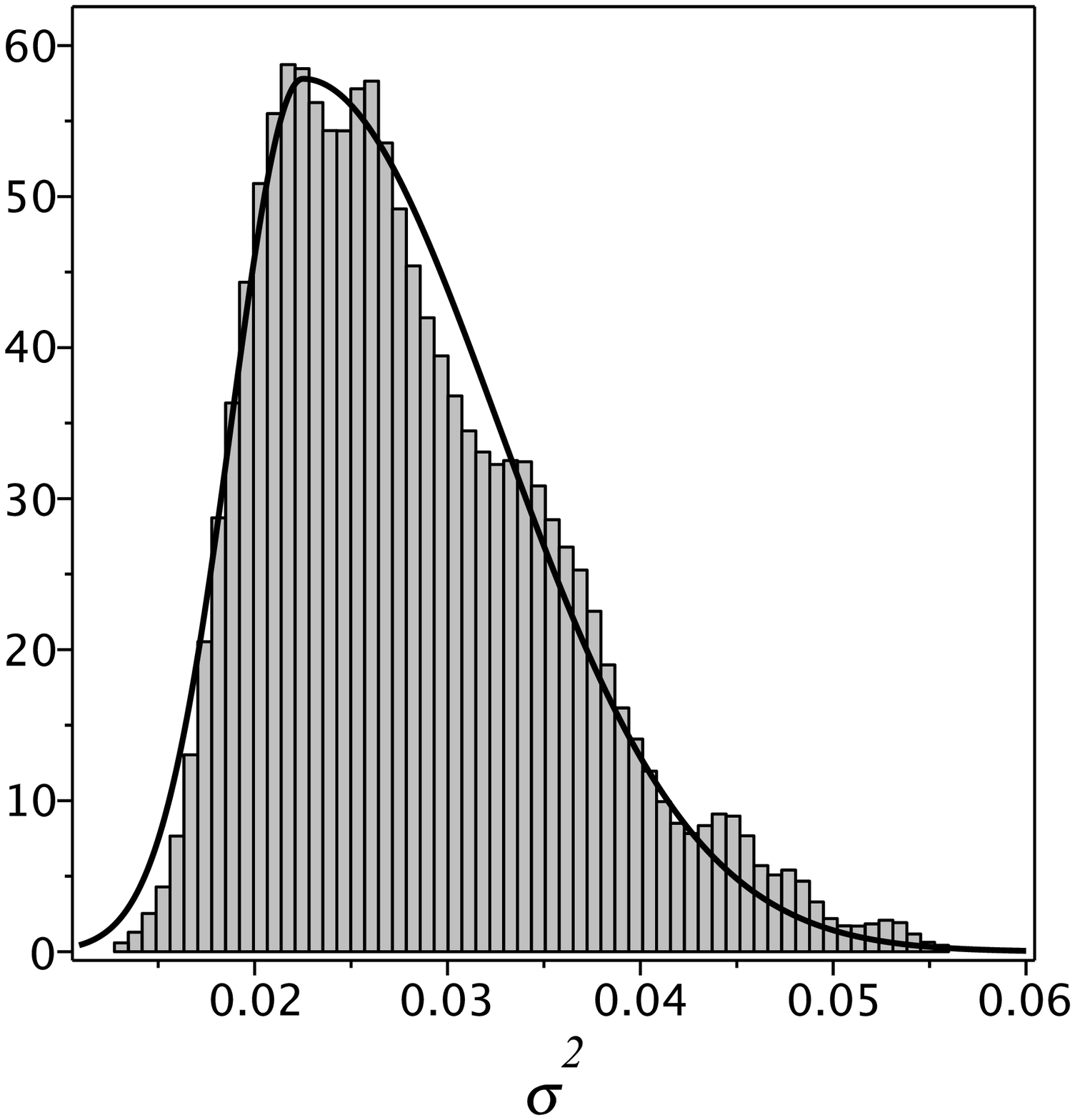}
\caption{Histograms of $\sigma^2(\tau)$ constructed with the same bin using intervals of time $\Delta \tau=100,300,600,800$ (from left to right and top to bottom), about $\tau=2,350$. It is clear that the same distribution is found no matter is the interval of time. This reproducible property of the turbulent signal indicates its self-similar nature.}
\end{figure*}

The aspect of wave turbulence relevant for the preheating is the cascade or transfer of energy among distinct scales. In particular, for preheating this is crucial to distribute the energy content from the homogeneous component of the inflaton to all perturbative modes. In this way, turbulence offers a very elegant mechanism to describe the transition from an empty and cold post-inflationary universe to a hot universe dominated by radiation in accordance with the hot big bang model. The imprint of such mechanism results in the so-called Kolmogorov-Zakharov spectrum \cite{nazarenko} that have the form of power-law in space and time domains. 

We start with the power spectrum of $\sigma^2(\tau)$ in the time domain shown in Fig. 4. The power spectrum presents a very rich structure. We have recognized three typical frequencies delimiting four regions with scaling laws. The first frequency is $\omega \approx 0.5656$, which is the smallest natural frequency associated to the modes with $|\mathbf{k}|=1$. The second frequency, $\omega \approx 0.8695$, is the initial frequency of the unperturbed homogeneous component of the inflaton, $a_\mathbf{0}(\tau)$, and the third frequency, $\omega \approx 15.99$, is the highest frequency associated to those modes with $|\mathbf{k}|_{max} = N \sqrt{2}=20\sqrt{2}$. It is necessary to comment about the first region of low frequencies $\omega \leq \omega_{\mathrm{min}}$. These frequencies originate from the period bifurcations which take place in route to turbulence \cite{ruelle,deol1,deol2}. In the intervals $\omega \leq \omega_{min}$ and $\omega \geq \omega_{max}$, we found that $P(\omega) \sim \omega^{-2.51}$ while for $\omega_{min} \leq \omega \leq \omega_{max}$ there are two regions for which $P(\omega) \sim \omega^{-6.52}$. These scaling laws in the time domain together with the scaling law in the space domain are a direct consequence of the turbulent process in the late stages of preheating. They are representative of a KZ spectrum for a steady-state turbulent system.

\begin{figure}[htb]
\centering
\includegraphics[height=6.5cm,width=8.5cm]{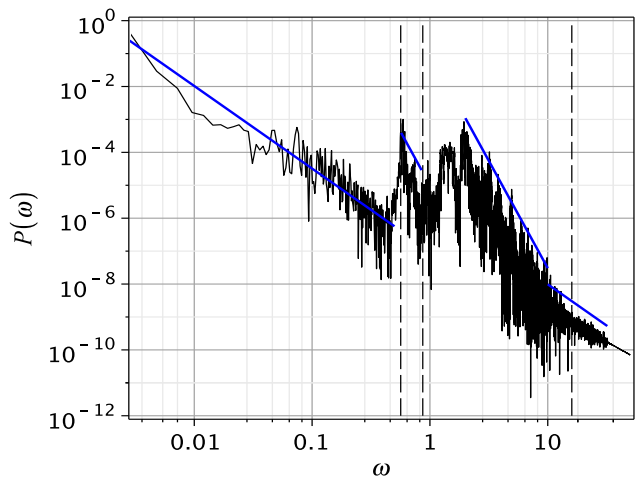}
\caption{Power spectrum in time domain of the spatial average of variance evaluated for $\lambda=10^{-4}$. The vertical lines indicate, from left to right three values of typical frequencies, $\omega \approx 0.5656, 0.8695$ and $15.99$. These are the minimum frequency, the initial frequency of the homogeneous inflaton field, and the maximum frequency, respectively. Here the minimum and maximum natural frequencies of the modes. It is clear the scaling law present in each region. The same structure of power spectrum is obtained for other values of $\lambda$ and larger intervals of times.}
\end{figure}

Besides the power spectrum in the time domain, we can calculate the power spectrum in the space domain of the variance. It is convenient to expand it with respect to the basis functions,

\begin{equation}
\mathrm{var}(\varphi) = (\varphi - \left<\varphi\right>)^2 = \sum_{\mathbf{k}}\,b_{\mathbf{k}}(\tau) \psi_{\mathbf{k}}(\mathit{{\mathbf{x}}}), \label{eq7}
\end{equation}    

\noindent where the modes $b_{{\mathbf{k}}}$ are associated to the wavenumber vector ${\mathbf{k}}$; in particular the $0th$-mode is  the spatial average of the variance, $b_{\mathbf{0}}=\sigma^2$. From the above expansion, we can construct the power spectrum in wave numbers $k=|\mathbf{k}|$ at several times. In Fig. 5 we show the power spectrum in the space domain evaluated at $\tau = 15,000$ with $\lambda = 10^{-8}$. The structure of the power spectrum almost does not change if calculated at any instant during the turbulent phase, in this case $\tau \geq 4,000$. We were able to fit the whole spectrum by a curve that contains both the power law and exponential factors, 

\begin{equation}
P(k) = 0.00689\,k^{-0.51}\mathrm{e}^{-0.0211 k^{1.54}}. \label{eq8}
\end{equation}

\noindent This kind of spectrum decay in wave numbers occurs in magnetohydrodynamic turbulence \cite{terry}. 

\begin{figure}[htb]
\centering
\includegraphics[height=6.5cm,width=8.5cm]{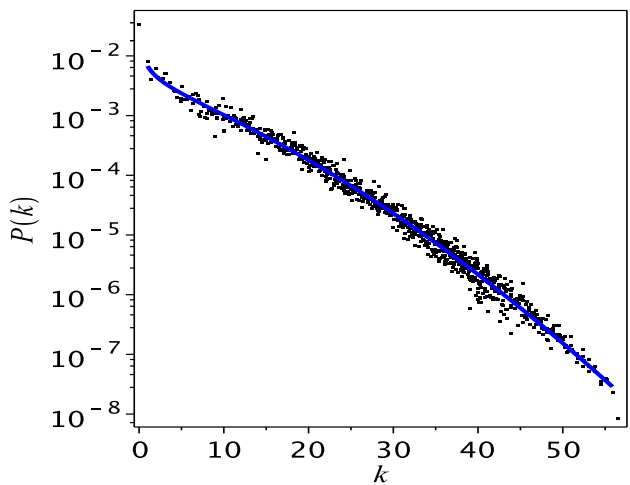}
\caption{Power spectrum $P(k)$ of the variance in the space domain evaluated at $\tau=15,000$ and with $\lambda=10^{-8}$. The continuous line described by Eq. (\ref{eq8}) fits the whole spectrum. A similar combination of exponential and power decay can be found in magnetohydrodynamic turbulence. We have considered $\lambda=10^{-8}$.}
\end{figure}

As established by the spectral approximation of Eq. (\ref{eq2}), the classical modes $a_{\mathbf{k}}=\alpha_{\mathbf{k}}+\beta_{\mathbf{k}}$ are interpreted as $c$-number amplitudes associated to processes of creation and annihilation of quantum fluctuations of the inflaton field in the mode $\mathbf{k}$. In this case, a relevant quantity is the occupation number of created particles, $n_k$, given by,

\begin{equation}
n_k = \frac{1}{\omega_{\mathbf{k}}} |\dot{a}_{\mathbf{k}}|^2+\frac{\omega_{\mathbf{k}}}{2}|a_{\mathbf{k}}|^2
\end{equation} 

\noindent where $\omega_{\mathbf{k}}=2 \pi |\mathbf{k}|/L$. We have considered $\lambda=10^{-8}$ and evaluated the power spectrum of $n_k$ with respect to $k=|\mathbf{k}|$ at $\tau = 15,0000$. The power spectrum showed in Fig. 6 was also found by Micha and Tkachev \cite{micha} using a different numerical approach in a 3D cubic lattice with a grid of $256^3$ points. Accordingly, the scaling law  $n(k) \sim k^{-s}$ for small $k$ (straight line) with $s \approx 1.69$, is in agreement of the Micha's result.  

The turbulent phase of preheating produces a thermalized universe. A possible way of evaluating the temperature of this thermalized phase is to calculate the power spectrum in the space domain of the energy density $a^4\rho_\phi(\mathbf{x},\tau)/\lambda \phi_e^4$ (cf. Eq. (\ref{eq14})), whose expansion with respect to the basis function is,

\begin{equation}
\frac{a^4\rho_\phi(\mathbf{x},\tau)}{\lambda \phi_e^4} = \sum_{\mathbf{k}}\,E_{\mathbf{k}}(\tau) \psi_{\mathbf{k}}({\mathbf{x}}).\label{eq9}
\end{equation}

\noindent With the above expansion, we have evaluated the spectrum of energy $\mathcal{E}(k)$ in wave numbers $k$, where the energy $\mathcal{E}(k)$ is the root mean squared of all energies $E_{\mathbf{k}}$ whose corresponding $\mathbf{k}$ has modulus $k$. In Fig. 7 we show the power spectrum at $\tau = 15,000$ and corresponding to $\lambda=10^{-8}$. The structure of the power spectrum displays two components separated by a gap of energy can also be found in Ref. \cite{deol2}. Most of the points lie in the second component that corresponds to the energy distribution for large wave-numbers, or small scales, which can be interpreted as the inertial range. 

We can understand the second piece of the spectrum as resulting from the energy transfer from the homogeneous mode to small scale modes as expected in turbulence. The distribution at large wave numbers might correspond to a thermalized system consistent with a distribution of incoherent radiation satisfying the Planck distribution given by,

\begin{equation}
\label{eq10}
{\cal{E}}_k = {\cal{E}}_0\,\frac{k^3}{\mathrm{e}^{b k}-1},
\end{equation}
 
\noindent where ${\cal{E}}_0$ and $b$ being constants. The best fit of (\ref{eq9}) represented by the continuous line has $b \approx 0.18$ and ${\cal{E}}_0 \approx 10^{-4.9}$. Another interesting feature is that a considerable part of the first region of the distribution can be fitted by the above distribution with the same value of $b$, but ${\cal{E}}_0 \approx 10^{-3.9}$ (cf. Fig. 7). Since the parameter $b$ depends on the temperature of the distribution, both pieces of the distribution are thermalized with the same temperature. 
 
\begin{figure}[htb]
\centering
\includegraphics[height=6.5cm,width=8.5cm]{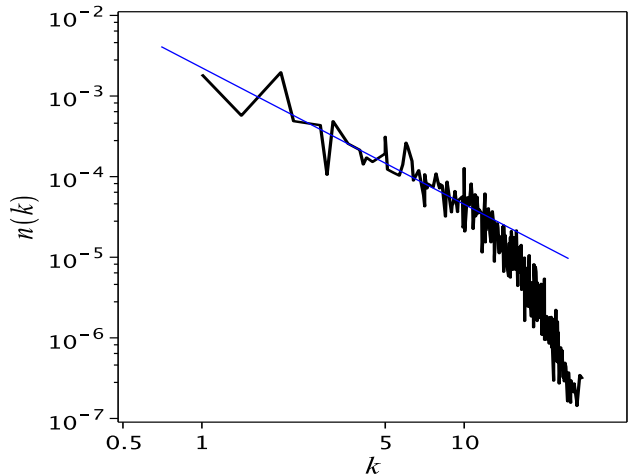}
\caption{Power spectrum of the occupation number $n(k)$. The straight line represents the scaling law $n(k) \sim k^{-1.69}$.}
\end{figure}
  
In order to extract the value of the temperature of the distribution, it is necessary to recover the physical variables present in the argument of the exponential from the dimensionless coordinates ${\bf x}$ and the momenta ${\bf k}$. In this case we have $b k = \hbar c k_{\mathrm{phys}}/k_B T$, where $k_{\mathrm{phys}}$ is physical momentum, $k_B$ and $T$ are the Boltzmann constant and temperature, respectively. From the relation $k =L k_{\mathrm{phys}}/\sqrt{\lambda}\phi_e a_0$, with $\phi_e = 0.1M_{pl},a_0 = 1$ \cite{inflation} corresponding to the scalar field, and scale factor at the end of inflation, the temperature of the distribution in natural units is estimated as,

\begin{equation}
\label{eq11}
T = \frac{\hbar c \sqrt{\lambda} \phi_e a_0}{k_B b L} \approx 10^{-5} M_{pl} \approx 10^{14} GeV. 
\end{equation}

\noindent This temperature is consistent with the beginning of the radiation era \cite{thermal}. 

\begin{figure}[htb]
\centering
\includegraphics[height=6.5cm,width=8.5cm]{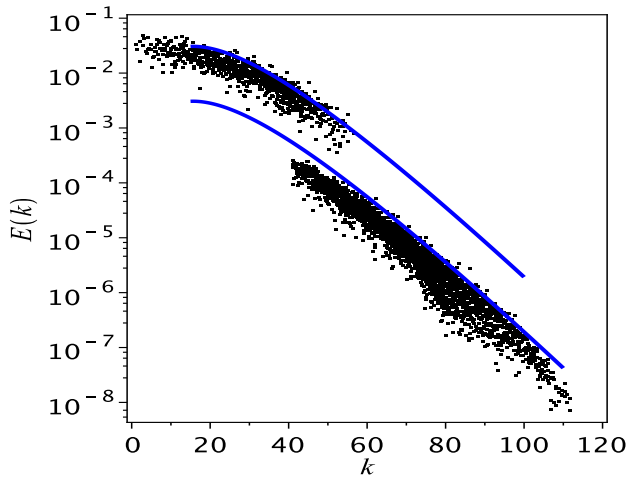}
\caption{Power spectrum $E(k)$ of the energy density of the inflaton field. Two components are present separated with a gap of energy. The continuous lines represent the Planck distribution with the same temperature, but different values of $\mathcal{E}_0$. This is an evidence that the whole system became thermalized. Here, the power spectrum was evaluated at at $\tau=15,000$ and $\lambda=10^{-8}$.}
\end{figure}

\section{Backreaction}

We have studied the evolution of the self-interacting inflaton field without taking into account the backreaction of its perturbative modes. The primary importance of the back-reaction is the generation of an effective energy-momentum equation. In the present model a perfect fluid emerges with an effective equation of state, 

\begin{equation}
w = \frac{\left<p_\phi\right>}{\left<\rho_\phi\right>}, \label{eq12}
\end{equation}

\noindent where $\left<..\right>$ represents the spatial averaging as mentioned before, and the pressure and energy density associated to the scalar field are, respectively,

\begin{eqnarray}
p_\phi = \frac{\lambda \phi_e^4}{a^4}\left[\frac{1}{2}\left(\varphi^{\prime}-\varphi \frac{a^\prime}{a}\right)^2 - \frac{1}{6} (\mathbf{\nabla} \varphi)^2 + \frac{1}{4} \varphi^4\right] \label{eq13}\\
\rho_\phi = \frac{\lambda \phi_e^4}{a^4}\left[\frac{1}{2}\left(\varphi^{\prime}-\varphi \frac{a^\prime}{a}\right)^2 + \frac{1}{2} (\mathbf{\nabla} \varphi)^2 + \frac{1}{4} \varphi^4\right].\label{eq14}
\end{eqnarray}

In order to provide a consistent framework for including the back-reaction, we have integrated the  following field equations,

\begin{eqnarray}
&& a^{\prime2}  = \frac{\phi_e^2}{3 M_{pl}^2}\left<\frac{1}{2}\left(\varphi^{\prime}-\varphi \frac{a^\prime}{a}\right)^2 + \frac{1}{2} (\mathbf{\nabla} \varphi)^2 + \frac{1}{4} \varphi^4\right> \nonumber \\
\label{eq16}\\
&& \varphi^{\prime\prime}-\nabla^2 \varphi-\frac{a^{\prime\prime}}{a}\varphi+{\varphi}^3=0.\label{eq17}
\end{eqnarray}  

\noindent We have set $\phi_e = 0.1 M_{pl}$, $a(0)=1$, $\lambda=10^{-4}$ and the initial conditions for the conformal scalar field were the same used in the previous Section. We have integrated the field equations in a squared box of size $L=5 \pi/\sqrt{2}$. At each time step the integrals resulting from the space average were calculated using Gauss quadrature formulae \cite{collocation}. In Fig. 8 we show the evolution of $w(t)$. Notice that at the initial phase of preheating the homogeneous modes dominates producing an effective equation of state of radiation, $w \approx 1/3$ as expected in the present model \cite{turner}. At late stages $w(t)$ seems to converge to an approximate fixed value $w \approx 0.06$.

\begin{figure}[htb]
\centering
\includegraphics[height=7.5cm,width=7.5cm]{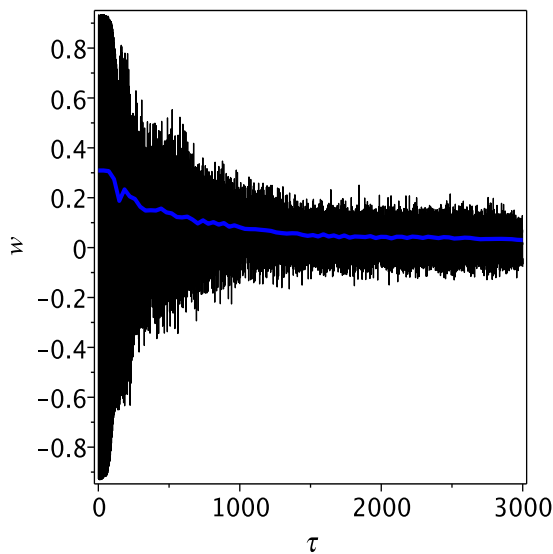}
\caption{Evolution of the effective equation of state parameter $w$ (blue line).}
\end{figure}

\section{Discussion}

In this paper we have studied the late stages of preheating of an inflationary model with potential $V(\phi)=\lambda \phi^4/4$. Although the present model is not in agreement, within an acceptable level of confidence, with current observational data, the stages towards the turbulence and thermalization seems to be robust and model independent.

We have evolved the field equation using collocation method. The complex modes $a_{\mathbf{k}}(\tau)$  in the spectral approximation of the inflaton field (Eq. (\ref{eq2}) can be viewed as the amplitudes of plane waves of wave number vector $\mathbf{k}$. Turbulence develops with the nonlinear interaction of these waves together with the transfer of energy from the homogeneous mode (zeroth mode) to other scales. The KZ spectra is an indicator of the energy transfer in a state of steady turbulence. Therefore, this mechanism plays a fundamental role in the establishment of thermalization allowing the universe to enter in the radiation era. We were able to determine the temperature of the thermalized state by assuming the spectrum of the energy density is consistent with the Plank distribution.

We intend to explore further the interplay between wave turbulence processes in the early universe. According to the latest observational data most one-field inflationary models becomes unfavorable, but we mention two one-field models worth of investigation. The first was proposed by Amin et al \cite{model_1} characterized by the potential,

\begin{equation}
V(\phi) = \frac{m^2 M^2}{\alpha} \left[\left(1+\frac{\phi^2}{M^2}\right)^\alpha-1\right].\label{eq18}
\end{equation}  

\noindent The authors have pointed out that choosing $\alpha < 1$ is in according with the observational data, and also it is possible to generate a preheating phase with broad resonance. The second model was proposed by Kallosh and Linde \cite{model_2}, and consists in a class of self interacting single field with the same potential treated in this work, but the field is non-minimally coupled with curvature \cite{linde}.

Finally, another relevant topic is the possibility of the parametric growth of gravitational perturbations during the preheating \cite{brand_finelli}. Therefore, it might be possible that the nonlinear phase undergoes what could be called gravitational wave turbulence.

\section*{Acknowledgements}

The authors acknowledge the financial support of the Brazilian agencies CNPq, CAPES and FAPERJ.

\end{document}